\documentclass[twocolumn,prb,amsmath,amssymb]{revtex4}
\usepackage{amsmath}
\usepackage{amssymb}

\usepackage{graphicx}

\begin{document}

\title{Hall effect anomaly near $T_{c}$ and renormalized superconducting fluctuations
in YBa$_{2}$Cu$_{3}$O$_{7-x}$}

\author{I. Puica}

\altaffiliation[Also at ]{the Department of Physics, Polytechnic
University of Bucharest, Spl. Independentei 313, RO-77206 Bucharest 6,
Romania}

\author{W. Lang}

\email{wolfgang.lang@univie.ac.at}

\author{W. G\"{o}b}

\affiliation{Institut f\"{u}r Materialphysik der Universit\"{a}t Wien, Boltzmanngasse
5, A-1090 Wien, Austria}

\author{Roman Sobolewski}

\altaffiliation[Also at ]{the Institute of Physics, Polish Academy of Sciences,
PL-02668
Warszawa, Poland}

\affiliation{Department of Electrical and Computer Engineering and Laboratory for Laser
Energetics, University of Rochester, Rochester, New York 14627-0231}

\date{\today{}}

\begin{abstract}
Measurements of the Hall effect and the resistivity on precisely-patterned
YBa$_{2}$Cu$_{3}$O$_{7-x}$ thin films in magnetic fields $B$ from 0.5
to 6 T oriented parallel to the sample crystallographic $c$ axis reveal
a sign reversal of the Hall coefficient for $B\leq 3$ T. The data are
compared to the full, quantitative expressions based on the renormalized
fluctuation model for the Hall conductivity. The model offers a satisfactory
understanding of the experimental results, for moderate fields and temperatures
near the critical region, provided that the inhomogeneity of the critical
temperature distribution is also taken into account. We also propose an
approach how vortex pinning that strongly affects the magnitude of the
Hall coefficient can be incorporated in the model.
\end{abstract}


\maketitle

\section{Introduction}

The influence of superconducting fluctuations on off-diagonal components
of the magnetoconductivity tensor (usually denoted as the excess Hall effect)
in high-temperature superconductors (HTSC) has received considerable experimental
and theoretical attention over the past few years.\cite{Wang,Vinokur93,Lang94,Samoilov,WuLiu,Roa01,Koku01,Kang1}
Though a general consensus seems to be achieved now regarding the existence
and the temperature dependence of the excess Hall effect, theoretical predictions
of its sign are still controversial. Experimentally, the Hall resistivity
shows a peculiar temperature dependence. Specifically, as the temperature
is decreased through the fluctuation region, the Hall resistivity decreases
and changes its sign relatively to the normal state one, exhibits a negative
minimum and eventually reaches zero at low temperatures. This simple sign
change was detected in many different HTSC\cite{Rice1,Clinton,Kang1,Matsuda,Roa01}
and even in conventional superconductors.\cite{Hagen,Graybeal,Koku01}
Furthermore, a double sign reversal, that is a subsequent return of the
Hall resistivity to the positive value before vanishing, has been observed
in highly anisotropic HTSC, such as Bi$_{2}$Sr$_{2}$CaCu$_{2}$O$_{x}$
crystals\cite{Samoilov1} and films,\cite{Zavaritsky} Tl$_{2}$Ba$_{2}$CaCu$_{2}$O$_{x}$
films,\cite{Hagen1} or HgBa$_{2}$CaCu$_{2}$O$_{6}$ films.\cite{Kang2}
Recently, the existence of the second sign change was also reported in
YBa$_{2}$Cu$_{3}$O$_{7-x}$ films, either at high current densities\cite{Nakao}
or in the strong pinning limit at low magnetic fields.\cite{Goeb00} Finally,
even a triple sign reversal was reported in HgBa$_{2}$CaCu$_{2}$O$_{6}$
films with columnar defects induced by high-density ion irradiation.\cite{Kang3}

Several theoretical approaches have attempted to explain the complex features
of the Hall resistivity temperature dependence, but no consensus has been
achieved. The Hall anomaly might originate from the pinning force,\cite{Wang}
non-uniform carrier density in the vortex core,\cite{Otterlo,Kato99} or
can be calculated in the time-dependent Ginzburg-Landau (TDGL) model.\cite{Troy,Kopnin93}
Most recent theories claim to predict the double or triple sign reversal,
based either on entirely intrinsic mechanism of vortex motion and electronic
spectrum,\cite{Kopnin96} or on hydrodynamic interaction between vortices
and the superconducting and normal state fluids.\cite{Kolacek} Some theories
invoke superconducting fluctuations alone to account for the Hall effect
sign reversal \cite{UD,NE}, while others present a more extended picture
based on the same foundations of TDGL using both the hydrodynamics and
the vortex charging effect, arising from the difference in electron density
between the core and the far outside region of the vortex.\cite{Otterlo,Feigelman,Kato99}
Thus, the Hall effect in the mixed state of HTSC reflects a complex interplay
between electronic properties of quasiparticles, thermodynamic fluctuations,
hydrodynamic effects of vortices, and pinning.

From a considerable part of the published theoretical work, it appears
that at least the first sign reversal, which occurs near the critical
region, where vortex pinning is negligible and the superconducting order
parameter fluctuations play an important role, should be ascribed to a
microscopic origin of
superconductivity.\cite{Samoilov,Smith,Graybeal,Beam} From the viewpoint
of the TDGL formalism,\cite{UD,Kopnin93,NE} to which any theory of vortex
dynamics must reduce near the critical temperature
$T_{c}$,\cite{Ikeda99,Kopnin96} the Hall anomaly is a consequence of the
difference in sign between the normal (quasiparticle) part and the
superconducting fluctuation (or vortex flow) part of the total Hall
conductivity. These two components have opposite signs, if the energy
derivative of the density of states averaged over the Fermi surface is
positive when the carriers are holes in the normal state.\cite{Kopnin95}
Thus, the sign reversal can be intrinsic and depends on the details of the
structure of the normal-state electronic spectrum. Such notion is further
supported by the fact that in several HTSC, the sign reversal disappears
when the material is strongly overdoped and the band structure approaches
that of a conventional metal.\cite{Nagaoka98}

The possibility of the Hall angle sign change in the critical region was
first discussed by Fukuyama, Ebisawa and Tsuzuki (FET),\cite{FET} who
pointed out that the origin of a non-vanishing Hall current due to fluctuating
Cooper pairs could come from a hole-particle asymmetry, which reveals a
complex relaxation time in the TDGL theory. In this early work, it was
implicitly assumed that the fluctuations did not interact; that is, only
Gaussian fluctuation were considered. Accordingly, the fluctuation parts
of the conductivity tensor elements were predicted to diverge at $T_{c}$
in the presence of magnetic field. However, this predicted divergence has
not been observed. A great improvement was obtained when the interaction
between fluctuations was taken into account by incorporating the quartic
term $\left|\psi \right|^{4}$ from the Ginzburg-Landau (GL) expression
of the free energy. Such a treatment was performed by Ullah and Dorsey\cite{UD}
(UD) in the frame of a simple Hartree approach of the TDGL theory. More
recently, Nishio and Ebisawa\cite{NE} (NE) extended the FET calculations
of the weak (Gaussian) fluctuation contribution of the Hall conductivity
to the strong (non-Gaussian) fluctuation regime, based on more sophisticated
renormalization theory by Ikeda, Ohmi and Tsuneto (IOT).\cite{IOT} The
renormalized, non-Gaussian fluctuation regime connects therefore the weak
(Gaussian) fluctuation regime in the paraconducting region above $T_{c2}\left(H\right)$
to the vortex liquid (flux-flow) regime below the mean-field transition,
interpolating smoothly without the $T_{c}$ divergence predicted by the
Gaussian theory.

In this paper, we present simultaneous measurements of the resistivity
and Hall resistivity, of epitaxial YBa$_{2}$Cu$_{3}$O$_{7-x}$ (YBCO)
films in a wide range of applied magnetic fields (from 0.5 to 6 T), and
give a quantitative account for our Hall-effect experimental data by using
the aforementioned renormalized fluctuation model of NE. \cite{NE} It
is worth mentioning that for $B<0.5$ T, we have earlier found an occurrence
of the second sign reversal\cite{Goeb00} in similar YBCO thin films in
$B$ fields oriented parallel to the crystallographic $c$ axis and to
the twin boundaries. This second sign change turned out to be strongly
vortex-pinning dependent, since it vanished at high transport current densities,
or with the $B$ field tilted off the twin boundaries by a small angle
($5^{\circ }$). For moderate magnetic fields instead, as those investigated
in the present paper, and for temperatures near the critical region, where
the first sign change occurs, the pinning contribution to the Hall conductivity
is almost negligible. \cite{Goeb00} The TDGL approach is therefore considered
to be appropriate, although quantitatively less accurate towards lower
temperatures and fields, where pinning becomes more effective.

There have been so far several reported verifications of merely scaling
relationships connecting fluctuation conductivities, temperature, and magnetic
field, emerging from the TDGL model. Liu \emph{et al.}\cite{WuLiu} found
good experimental evidence for the validity of the scaling laws depending
on temperature and $B$ field given by the Hartree renormalization procedure
in the lowest Landau level.\cite{UD} This approach was applied to the
Aslamazov-Larkin (AL) term of the fluctuation longitudinal and Hall conductivities,\cite{AL}
for $B$ fields ranging between $2$ and $9$ T, and identified the cause
of the Hall conductivity sign change as lying in the fluctuation regime.
Ginsberg and Manson\cite{Ginsberg} and Neiman \emph{et al.},\cite{Neiman}
also found a satisfactory fit for their data by using the $1/B$ proportionality
of the fluctuation Hall conductivity, predicted by the same Hartree renormalization
fluctuation model\cite{UD} in the lowest Landau level approximation (valid
in the high field limit). We have, however, no knowledge of any direct
comparison between experimentally observed Hall anomaly in HTSC and the
full quantitative application of the TDGL theory in the renormalized fluctuation
regime, where the first sign reversal occurs. And this is the main purpose
of this work. In Sec. \ref{theory}, the most important results of the
IOT\cite{IOT} and NE\cite{NE} renormalized fluctuation theories for the
longitudinal and Hall conductivities are reviewed, and modifications for
including samples with nonuniform $T_{c}$'s are proposed. Section \ref{exp}
briefly presents our sample preparation and measurement techniques, while
Sec. \ref{results} shows our experimental results and directly compares
them to the theoretical model. Finally, in Sec. \ref{conclu}, we summarize
our results and list the main conclusions emerging from our analysis.

\section{Theoretical background}

\label{theory}Based on the IOT renormalization theory, NE extended FET
calculations of the weak fluctuation contribution of Hall conductivity
and derived an expression of the excess Hall conductivity $\Delta \sigma _{xy}$
due to the non-Gaussian superconducting fluctuations corresponding to the
AL process in a layered superconductor:

\begin{eqnarray}
 &  & \Delta \sigma _{xy}=\beta \frac{e^{2}h^{3}}{\hbar \xi _{c}}\frac{k_{B}T}{\varepsilon _{F}}\sum _{n=0}^{\infty }\frac{\left(n+1\right)}{\left(\varepsilon _{n+1}-\varepsilon _{n}\right)^{2}}\nonumber \\
 &  & \times \left[1+d^{2}\left(\varepsilon _{n}+\varepsilon _{n+1}\right)\right]\left(\frac{f_{n}^{2}f_{n+1}^{2}}{f_{n}+f_{n+1}}-\frac{1}{2}f_{n+\frac{1}{2}}^{3}\right)\: ,\label{1}
\end{eqnarray}
 with: $f_{n}=\left[\varepsilon _{n}\left(1+d^{2}\varepsilon _{n}\right)\right]^{-1/2}$;
$\varepsilon _{n}=\varepsilon _{0}+2nh$, $\left(n\geq 1\right)$; $\varepsilon _{0}=\varepsilon +h$;
$\varepsilon =\left(T-T_{c}\right)/T_{c}$; $h=2\pi \xi _{ab}^{2}B/\Phi _{0}$;
$d=s/2\xi _{c}$; $\beta =-4\varepsilon _{F}N'/\pi gN^{2}$. Here $N$
is the density of states at the Fermi surface $\varepsilon _{F}$, $N'$
is the energy derivative of $N$, $g$ ($>0$) the BCS coupling constant,
$\xi _{ab}$ and $\xi _{c}$ are the coherence lengths extrapolated at
$T=0$ in $ab$ and $c$ directions, respectively, $s$ is the distance
between superconductor layers in the Lawrence-Doniach\cite{LD} model,
$T_{c}$ is the critical temperature in the absence of the magnetic field,
and $B$ the magnetic field applied perpendicularly to the $ab$ plane.
The renormalization procedure, described in detail by IOT, consists in
using the renormalized expression $\tilde{\varepsilon }_{n}$ instead of
$\varepsilon _{n}$ for each Landau level $n$,

\begin{eqnarray}
 &  & \tilde{\varepsilon }_{n}=\varepsilon _{n}+\frac{g_{3}\, d}{\sqrt{\beta _{0}^{2}-1}}+\frac{\sqrt{\beta _{0}^{2}-1}}{8\beta _{0}\, d^{2}\left(n+1\right)\textrm{!}}\nonumber \\
 &  & \times \left\{ \left(\ln \frac{\gamma _{+}}{\alpha _{+}}\right)^{n+1}+\frac{\alpha -\beta _{0}}{\sqrt{\beta _{0}^{2}-1}}\right.\nonumber \\
 &  & \times \left.\left[\ln \left(\frac{\beta _{0}\gamma +\sqrt{\left(\beta _{0}^{2}-1\right)\left(\gamma ^{2}-1\right)}-1}{\beta _{0}\alpha +\sqrt{\left(\beta _{0}^{2}-1\right)\left(\alpha ^{2}-1\right)}-1}\right)\right]^{n+1}\right\} \: ,\label{2}
\end{eqnarray}
 where: $g_{3}=8\pi ^{2}\mu _{0}\kappa ^{2}k_{B}T_{c}\xi _{ab}^{4}B/\xi _{c}\Phi _{0}^{3}$;
$\beta _{0}=1+2d^{2}\tilde{\varepsilon }_{0}$; $\alpha =2\beta _{0}^{2}-1$;
$\gamma =\alpha +8g_{3}\beta _{0}d^{3}(\beta _{0}^{2}-1)^{-1/2}$; $\alpha _{+}=\alpha +\sqrt{\alpha ^{2}-1}$;
$\gamma _{+}=\gamma +\sqrt{\gamma ^{2}-1}$, and $\kappa $ being the GL
parameter of the superconductor. The second term on the right hand side
of Eq. (\ref{2}) is the Hartree term and always dominates the third one.

The pre-factor in Eq. (\ref{1}) can be modified in a form more convenient
for experimental data fits. Taking into account the coherence length expression
valid in the dirty limit:\cite{Larkin02} $\xi _{ab}=\left(\pi \hbar \, v_{F}l/24k_{B}T_{c}\right)^{1/2}$
(where $v_{F}$ is the Fermi velocity, $\tau $ the scattering time and
$l=v_{F}\tau $ is the mean free path) together with the normal-state Hall
conductivity expression in the classical picture:\cite{FET} $\sigma _{xy}^{\mathrm{n}}=\sigma _{xx}^{\mathrm{n}}eB\tau /m_{e}$,
one can obtain the following form for $\Delta \sigma _{xy}$ when $T\approx T_{c}$:

\begin{eqnarray}
 &  & \Delta \sigma _{xy}=\beta \frac{\pi \, e^{2}}{24\hbar \xi _{c}}\cdot \frac{\sigma _{xy}^{\mathrm{n}}}{\sigma _{xx}^{\mathrm{n}}}\cdot \sum _{n=0}^{\infty }\frac{4h^{2}\, \left(n+1\right)}{\left(\tilde{\varepsilon }_{n+1}-\tilde{\varepsilon }_{n}\right)^{2}}\nonumber \\
 &  & \times \left[1+d^{2}\left(\tilde{\varepsilon }_{n}+\tilde{\varepsilon }_{n+1}\right)\right]\left(\frac{\tilde{f}_{n}^{2}\tilde{f}_{n+1}^{2}}{\tilde{f}_{n}+\tilde{f}_{n+1}}-\frac{1}{2}\tilde{f}_{n+\frac{1}{2}}^{3}\right)\: .\label{3}
\end{eqnarray}
 with $\tilde{f}_{n}=\left[\tilde{\varepsilon }_{n}\left(1+d^{2}\tilde{\varepsilon }_{n}\right)\right]^{-1/2}$
and $\tilde{\varepsilon }_{n}$ given by Eq. (\ref{2}). Expression (\ref{3})
concerns, as stated, a layered superconductor, and only the pre-factor
was computed assuming a three-dimensional (3D) isotropic Fermi surface,
as in the BCS theory, justified by the moderate YBCO anisotropy. Considering
instead a cylindrical Fermi surface, corresponding to the two-dimensional
(2D) case, would change the $\xi _{ab}$ expression by a factor of $\sqrt{3/2}$,\cite{Larkin02}
and consequently the $\beta $ value by a factor of $2/3$. Since the correct
band structure for YBCO is expected to be in between these two limit cases,
our estimation for $\beta $ will be only slightly affected. Noticing that
$\tilde{\varepsilon }_{n+1}-\tilde{\varepsilon }_{n}\approx 2h$, one can
verify that the above formula gives in the low-field limit ($h\ll \varepsilon $)
in the paraconducting region (above $T_{c}$), an expression that formally
matches the 2D and 3D results of the FET theory for the AL fluctuation
term. The 2D limit corresponds to $\varepsilon d^{2}\gg 1$, while the
3D limit is valid when $\varepsilon d^{2}\ll 1$. The essential difference
remains, however, the presence in Eq. (\ref{3}) of the $\varepsilon _{n}$-renormalization,
according to the IOT theory of non-Gaussian superconducting fluctuations.

The IOT theory also gives the fluctuation contribution to the longitudinal
conductivity in the renormalized regime:

\begin{equation}
\Delta \sigma _{xx}=\frac{e^{2}h^{2}}{2\hbar \xi _{c}}\sum _{n=0}^{\infty }\frac{n+1}{\left(\tilde{\varepsilon }_{n+1}-\tilde{\varepsilon }_{n}\right)^{2}}\left(\tilde{f}_{n}+\tilde{f}_{n+1}-2\tilde{f}_{n+\frac{1}{2}}\right)\: ,\label{4}\end{equation}
 which, as derived from the GL functional, corresponds to the AL process.
It is worth mentioning that the sums over Landau levels in Eqs. (\ref{1})
and (\ref{4}) given by the IOT - NE theory correspond formally to those
found in the results of UD\cite{UD} in the frame of a simple Hartree approach
for incorporating the $\left|\psi \right|^{4}$ term in the TDGL theory,
with the specification that the UD renormalization procedure retains only
the Hartree contribution.

In previous papers,\cite{Lang,Heine} it was reported that even minor inhomogeneities
of $T_{c}$ within the sample may have a measurable, quantitative effect
to the paraconductivity, fluctuation magnetoconductivity, and excess Hall
conductivity. Thus, we are going to take this effect into account in our
derivations. Retaining only the first order expansion term of the effective
medium approximation, the inhomogeneity correction writes simply as an
average of fluctuation conductivities over the $T_{c}$ distribution. For
simplicity, we assume in our analysis a Gaussian distribution of $T_{c}$'s
with a mean value $T_{c0}$ and a standard deviation $\delta T_{c}\ll T_{c0}$,
so that the mean fluctuation conductivities will write as: \begin{equation}
\left\langle \Delta \sigma \right\rangle =\int \frac{1}{\delta T_{c}\sqrt{2\pi }}\exp \left[-\frac{\left(T_{c}-T_{c0}\right)^{2}}{2\left(\delta T_{c}\right)^{2}}\right]\cdot \Delta \sigma \left(T_{c}\right)\; \mathrm{d}T_{c}\: ,\label{5}\end{equation}
 where $\Delta \sigma $ stands for both $\Delta \sigma _{xy}$ and $\Delta \sigma _{xx}$.

The averaged fluctuation conductivities $\left\langle \Delta \sigma _{xy}\right\rangle $
and $\left\langle \Delta \sigma _{xx}\right\rangle $ derived from Eqs.
(\ref{3}), (\ref{4}) and (\ref{5}) have to be added to the normal state
components, $\sigma _{xy}^{\mathrm{n}}$ and $\sigma _{xx}^{\mathrm{n}}$,
respectively. No general consensus exists about the functional form of
$\sigma _{xx}^{\mathrm{n}}$ for HTSC, but the linear temperature dependence
of resistivity over a broad temperature range is generally accepted. It
has also been shown\cite{Harris} that many Hall effect measurements in
various HTSC materials can be explained using the Anderson's formula:\cite{Anderson}
$\cot \theta _{H}^{\mathrm{n}}=\sigma _{xx}^{\mathrm{n}}/\sigma _{xy}^{\mathrm{n}}=C_{1}T^{2}+C_{0}$.
We shall therefore use for the normal-state part of the conductivity tensor
the simple expressions: \begin{equation}
\sigma _{xx}^{\mathrm{n}}=\frac{1}{p_{0}+p_{1}T}\: \: \: \text {and}\: \: \: \sigma _{xy}^{\mathrm{n}}=\frac{1}{p_{0}+p_{1}T}\cdot \frac{1}{C_{1}T^{2}+C_{0}}\: ,\label{6}\end{equation}
 where $p_{0}$, $p_{1}$, $C_{1}$ and $C_{0}$ are fitting parameters
to be determined from the experiment. Thus, the full conductivities will
be consequently:

\begin{equation}
\sigma _{xx}=\sigma _{xx}^{\mathrm{n}}+\left\langle \Delta \sigma _{xx}\right\rangle \: \: \text {and}\: \: \sigma _{xy}=\sigma _{xy}^{\mathrm{n}}+\left\langle \Delta \sigma _{xy}\right\rangle \: ,\label{7}\end{equation}
 where $\left\langle \Delta \sigma _{xy}\right\rangle $ is given by Eqs.
(\ref{3}) and (\ref{5}), $\left\langle \Delta \sigma _{xx}\right\rangle $
by Eqs. (\ref{4}) and (\ref{5}), and $\sigma _{xy}^{\mathrm{n}}$ and
$\sigma _{xx}^{\mathrm{n}}$ by Eq. (\ref{6}).

In the above considerations, we only included the AL process as contribution
to the fluctuation Hall and longitudinal conductivities. In order not to
overcomplicate the model by introducing non-essential parameters, we neglect
the Maki-Thompson\cite{Maki} and the density-of-states terms,\cite{Varlamov99}
which contribute as corrections to the excess conductivities only when
$B\ll 1$ T, and only in the above-$T_{c}$ region. They give therefore
a negligible contribution to the sign change features of the Hall resistivity,
and, moreover, their influence can hardly be quantitatively discerned from
a small correction of the normal-state fit.

\section{Experimental techniques}

\label{exp} Our epitaxial YBCO thin films were deposited by single-target
rf sputtering on $\textrm{LaAlO}_{3}$ substrates and patterned to precisely
aligned test structures using a laser inhibition technique on a computer-controlled
xy-stage.\cite{Kula95} Electrical contacts were established with gold
wires, attached by silver paste to evaporated silver pads. The onset of
the superconducting transition in zero field was at 90 K and the critical
current density of our films exceeded 3 MA/cm$^{2}$ at 77 K.

The experiments were performed with 17-Hz ac currents at $j=250$ A/cm$^{2}$
together with lock-in detection. The measurements from 1 to 6 T were made
in a commercial superconducting solenoid, while low-field measurements
at 0.5 and 1 T were performed in a closed-cycle cryocooler and with an
electromagnet. Results obtained from these two different set-ups were identical
at $B=1$ T. More detailed description of our experimental systems can
be found in Refs. \onlinecite{Goeb00} and \onlinecite{Kula95}.

\section{Results and discussions}

\label{results}
\begin{figure}
\includegraphics[  height=2.4in]{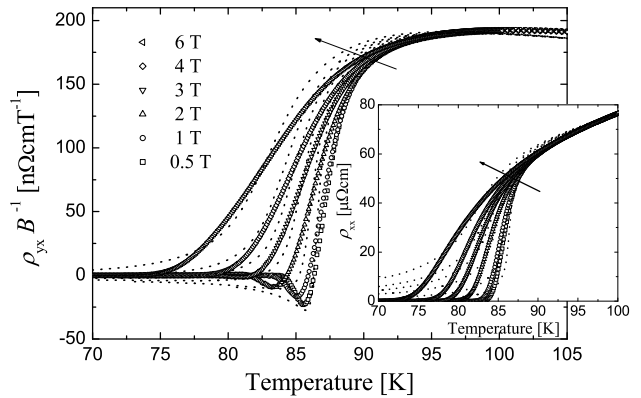}

\caption{\label{Fig1} Comparison between the experiment (symbols) and the NE renormalized
fluctuation model (dotted lines) for the YBCO normalized Hall resistivity
$\rho _{yx}/B$ as a function of temperature for different values of the
magnetic field. The inset shows the comparison between the experiment (symbols)
and the IOT renormalized fluctuation model (dotted lines) for the YBCO
longitudinal resistivity $\rho _{xx}$. The arrows indicate the increasing
field direction. The fit parameters are given in the text.}
\end{figure}

The experimental Hall resistivity normalized to the field
for a YBCO thin film, measured in various magnetic fields is shown in Fig.
\ref{Fig1} (symbols), while in the inset in Fig. \ref{Fig1} the longitudinal
resistivity is presented (symbols). The superconducting transition in Fig.
\ref{Fig1} inset is typical for a thin-film sample with a vortex-glass
behavior at low temperatures, while the shape of the upper part of the
transition is common to both thin films and single crystals.\cite{Sarti}
Figure \ref{Fig1} shows that the Hall resistivity is always positive (hole-like)
for $B>3$ T and exhibits the sign change at lower fields, in accordance
with previous investigations performed in the similar magnetic field range.
One can also notice that the Hall resistivity minimum occurs in the vortex-liquid
regime, and that the Hall anomaly increases significantly when the field
is reduced below 2 T.

Our first attempt to fit the experimental data by using theoretical dependencies
of the renormalized fluctuation model\cite{NE,UD} is also shown in Fig.
\ref{Fig1} (dotted lines) and results in curves similar to those presented
in the IOT and NE theoretical papers. The effect of the sample inhomogeneity
was in this approach neglected. Eq. (\ref{4}), given by the IOT model,
was used for the fluctuation longitudinal conductivity, while Eq. (\ref{3}),
based on the NE model, provided the fluctuation Hall conductivity. Both
models rely on the same set of adjustable parameters. The normal state
contributions, obtained by fitting the experimental data for temperatures
greater than 100 K with the expressions given in Eq. (\ref{6}), were,
subsequently, added to the fluctuation contributions. Finally, the inverted
conductivity tensor gave the longitudinal and Hall conductivity shown in
Fig. \ref{Fig1}. The model parameters that allowed to find the best fitting
theoretical curves, were: $T_{c}=87$ K, $\kappa =70$, $s=1.17$ nm equal
to the $c$-axis lattice parameter (this implies that the two copper-oxide
planes in the unit cell are tightly coupled, acting as one superconducting
layer), $\xi _{ab}=1.2$ nm and $\xi _{c}=0.14$ nm at $T=0$, and $\beta =-0.007$.
Comparison between the experiment and the model in Fig. \ref{Fig1} shows
that the renormalized fluctuation approach is adequate, at least from a
qualitative point of view. All features of the Hall resistivity dependence
on temperature, namely the steep decrease in the fluctuation region below
90 K, the sign change, the negative minimum and subsequently the vanishing
trend at low temperatures are clearly reproduced by the model.

We note that the fitting parameters listed above, like the coherence lengths
and the Ginzburg-Landau parameter assume the values very typical for YBCO.
Essential for this approach is, however, the negative value of the hole-particle
asymmetry parameter $\beta $ (this means a negative $\Delta \sigma _{xy}$)
that implies a positive energy derivative of the density of states at $\varepsilon _{F}$
when the carriers are holes in the normal state. As suggested by Kopnin
and Vinokur,\cite{Kopnin99} one possibility to explain this behaviour
is that the Fermi surface of a metal in the normal state has both hole-like
and electronic pockets. The Hall anomaly may thus depend on the doping
level, as it was reported by Nagaoka \emph{et al}.\cite{Nagaoka98} Very
recently, Angilella \emph{et al.}\cite{Angilella03} have found that, close
to an electronic topological transition of the Fermi surface, in the hole-like
doping range, the fluctuation Hall conductivity has indeed an opposite
sign with respect to the normal state one, giving additional strong support
that the Hall resistivity sign reversal is intrinsic and depends on the
details of the structure of the electronic spectrum.

We shall further discuss the reasons for quantitative discrepancies between
experimental curves and the model predictions in Fig. \ref{Fig1}, and
provide some modalities to improve them. A first point is that the IOT
model for the longitudinal resistivity fails to reproduce correctly the
low temperature part of the transition, giving too long tail of the resistivity
decrease. Two reasons are responsible for this behavior. One of them lies
in the renormalization procedure in the IOT model, which roughly corresponds
to a Hartree approximation. As remarked by Ullah and Dorsey,\cite{UD}
an important consequence of the Hartree approximation is that the calculated
properties in the flux-flow limit differ from the mean-field predictions
by a numerical factor of $2/\beta _{A}$, where $\beta _{A}$ is the Abrikosov
parameter $\beta _{A}=\left\langle \left|\psi \right|^{4}\right\rangle /\left\langle \left|\psi \right|^{2}\right\rangle ^{2}=1.16$
for a triangular vortex lattice. Thus, the Hartree prediction for the conductivity
is $2/\beta _{A}$ times smaller than the mean field result, which consequently
leads to a higher resistivity predicted by the fluctuation model in the
low-temperature range of the transition, as it can be seen in the inset
in Fig. \ref{Fig1}. Another reason, maybe more important quantitatively,
is the presence of flux pinning, which is neglected in the fluctuation
model, but which drastically steeps the resistivity descent in the lower
part of the transition. This quantitative inadequacy of the fluctuation
model for the $\rho _{xx}$ in the low temperature region manifests itself
implicitly in the corresponding features of the $\rho _{yx}$ theoretical
curve, namely in the long, nonvanishing tail at low temperatures. For this
reason, we decided to further test the general validity of the renormalized
fluctuation model over the entire temperature range only for the Hall conductivity
$\sigma _{xy}=\rho _{yx}/\left(\rho _{xx}^{2}+\rho _{yx}^{2}\right)$ that
is believed to be almost independent of pinning.\cite{Vinokur93}

\begin{figure}
\includegraphics[  height=2.6in]{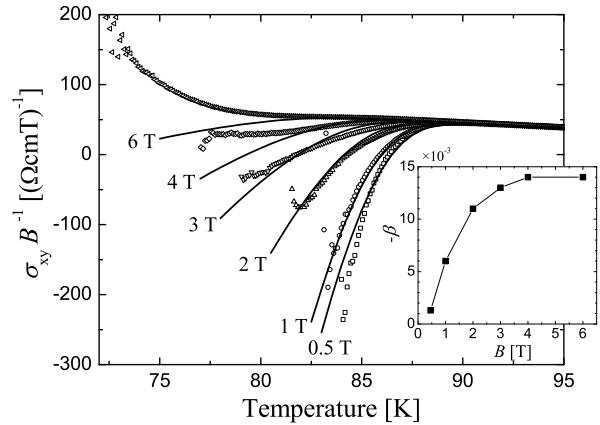}

\caption{\label{Fig2}Comparison between the experiment (symbols) and the renormalized
fluctuation model (solid lines) for the YBCO normalized Hall conductivity
$\sigma _{xy}/B$ as a function of temperature for different values of
the magnetic field. The NE model was used here with the relaxed $\beta $
parameter, in order to obtain the best fit. An average of the Hall conductivities
over a Gaussian distribution of $T_{c}$'s within the sample was also included.
The inset shows the $\beta $ dependence on the magnetic field, extracted
from fitting. The line is to guide the eye.}
\end{figure}

Following the above conclusion, Fig. \ref{Fig2} (symbols) presents the
experimental Hall conductivity $\sigma _{xy}$ normalized to $B$. It is
instructive to visualize the Hall effect using this plot, since $\sigma _{xy}/B$
is independent of $B$ in the normal state above 90 K. The observed behavior
suggests the presence of at least two contributions to the Hall conductivity.
One has the same sign as the normal-state effect and rapidly increases
below $T_{c}(B)$, becoming predominant for $B>4$ T, and indicating a
reduced carrier scattering in the superconducting state. The other contribution
exhibits an opposite sign and gains importance with smaller $B$'s. Thus,
for fields smaller than 3 T, the negative part dominates and $\sigma _{xy}$
changes its sign. It can be also noticed in Fig. \ref{Fig2} that with
decreasing $B$, the negative contribution shifts to higher temperatures
and exists in a narrower temperature range.

In small $B$ fields, the Hall anomaly is a very sharp feature in the experimental
data. A possible inhomogeneity of the material will influence the low-field
results, but remain insignificant at higher fields. In order to improve
the quantitative agreement with the experiment, we have included in our
model a distribution of $T_{c}$'s over the sample {[}see Eq. (\ref{5}){]},
and Eq. (\ref{7}) was used for the Hall conductivity. The main effects
of this correction are a less steep decrease of the Hall resistivity in
the first part of the transition (immediately below 90 K) and a relative
reduction in absolute value of the negative minimum. Figure \ref{Fig2}
presents the results of such a model (solid lines), where a Gaussian distribution
of $T_{c}$'s was used with a relative variance $\delta T_{c}/T_{c0}=0.02$.
All the other parameters except $\beta $, namely $\xi _{ab}$, $\xi _{c}$,
$s$ and $\kappa $ remained the same, as used in the fits shown in Fig.
\ref{Fig1}. We found that the best fits were obtained with a relaxed $\beta $
parameter, and inferred empirically an apparent field dependence of this
parameter, shown in the inset in Fig. \ref{Fig2}. We think, however that
the decrease of the $\beta $ absolute value with decreasing $B$ could
be simply the dissimulated effect of the increasing role of vortex pinning
at lower field values. Indeed, in our recent paper,\cite{Goeb00} a second
sign reversal was clearly identified for fields under 0.5 T and this effect
became more important with the decreasing field. The second sign change
disappeared in high current densities or under slightly tilted field direction,
revealing its vortex pinning origin. The positive pinning contribution
to the Hall conductivity, which gains importance at low fields could be
therefore reflected in the apparent field-dependence of the absolute value
of $\beta $. In a recent theoretical work, Kopnin and Vinokur\cite{Kopnin99}
also signalized, based on a simple model of pinning potential, that an
increasing pinning strength not only affected the longitudinal flux-flow
resistivity, but also decreased the magnitude of the vortex contribution
to the Hall voltage (fluctuation term in the TDGL approach). Strong enough
pinning can even result in a second sign reversal of the Hall resistivity,
if the negative vortex (fluctuation) contribution is reduced in absolute
value to magnitudes that are insufficient to counteract the positive contribution
of the normal state conduction.\cite{Kopnin99,Ikeda99}

\begin{figure}
\includegraphics[  height=2.4in]{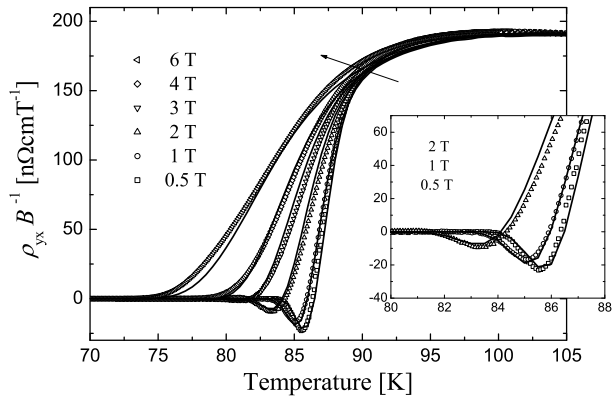}

\caption{\label{Fig3} Comparison between the experiment (symbols) and the renormalized
fluctuation model (solid lines) for the YBCO normalized Hall resistivity
$\rho _{yx}/B$ as a function of temperature for different values of the
magnetic field (the arrow indicates the increasing $B$ direction). The
experimental longitudinal conductivity was used for the calculation of
$\rho _{yx}$. The inset shows the transition temperature region in detail.
For further details see the text.}
\end{figure}

The best illustration of our modelling approach is presented in Fig. \ref{Fig3}
where we show the normalized Hall resistivity $\rho _{yx}/B$ computed
by using the \emph{theoretical} Hall conductivity $\sigma _{xy}=\sigma _{xy}^{\textrm{n}}+\left\langle \Delta \sigma _{xy}\right\rangle $
and the \emph{experimental} longitudinal conductivity $\sigma _{xx}^{\textrm{exp}}\cong 1/\rho _{xx}^{\textrm{exp}}$,
with the $T_{c}$-distribution correction included. The idea behind this
procedure is that the effect of pinning manifests itself primarily in $\sigma _{xx}$,
whereas $\sigma _{xy}$ is almost independent of pinning for magnetic fields
$\geq 1$ T, as it was shown in a number of different experiments using
artificially introduced defects \cite{Samoilov,Beam} or variation of current
density.\cite{Smith} We note an extremely good agreement between the experimental
and theoretical curves of the Hall resistivity in the entire temperature
and magnetic field ranges. For fields above 3 T, for which the transition
width enlarges towards lower temperatures (below 80 K), an additional positive
contribution to the Hall conductivity (see Fig. \ref{Fig2}), other than
the extrapolation of the normal state one, turns out to play a prevalent
role. This effect is now not evidenced in the Hall resistivity picture
(Fig. \ref{Fig3}), due to the vanishing Hall resistivity in this temperature
region. A possible interpretation for this supplementary positive term
to the Hall conductivity is the modification of the normal-state conduction
itself, namely, a reduced carrier scattering of quasiparticles in the superconducting
state. Figure \ref{Fig3} also proves that the much slower asymptotic trend
of the theoretical Hall resistivity towards zero observed in Fig. \ref{Fig1},
was indeed caused by the non-adequacy of the fluctuation model to the low-temperature
part of the longitudinal resistivity dependence. An improvement of the
model should therefore take into account also flux pinning, since it affects
the longitudinal conductivity in the lower part of the transition. It can
also be seen comparing Figs. \ref{Fig2} and \ref{Fig3} that including
the $T_{c}$ distribution results in smoothing of the curves and leads
to a gentler slope of the Hall resistivity in its initial positive part.
Still non-elucidated remains the true value of the $\beta $ parameter,
since the values returned by the fitting procedure are most likely altered
by the pinning effect on the Hall conduction and appear to be rather sensitive
to sample $T_{c}$ inhomogeneities. The hole-particle parameter was also
deduced from an independent analysis of the excess Hall effect caused by
Gaussian superconducting fluctuations above the mean-field critical temperature.\cite{Lang94}
Although the negative sign was found as well, the magnitude of $\beta $
differed significantly, likely due to the different limits in the models
on which the analysis was based. The negative sign of $\beta $, connected
with a positive derivative of the density of states at the Fermi level
is, however, essential in order to explain the sign change, from positive
(hole-like) to negative (electron-like) in the Hall resistivity.

\section{Conclusions}

\label{conclu} We presented results of simultaneous measurements of the
resistivity and Hall resistivity for epitaxial YBa$_{2}$Cu$_{3}$O$_{7-x}$
films in a wide range of the magnetic field, and explained our Hall-effect
experimental data by comparing them to the full quantitative expressions
given by the renormalized fluctuation model for the excess Hall conductivity
in HTSC materials. We found that this model offers an adequate quantitative
understanding of the experimental dependencies for moderate fields and
temperatures near the critical region, provided that the inhomogeneity
of the $T_{c}$ distribution is also taken into account. The essential
factor that explains the Hall anomaly is the negative fluctuation term
in the Hall conductivity, due in turn to the negative hole-particle asymmetry
parameter. In this framework, the Hall resistivity sign change and the
presence of the negative minimum for magnetic fields lower than 3 T is
easily accounted for. We have also found that for high fields ($B\geq 4$
T), in the lowest temperature range of the transition, the positive contribution
to the Hall conductivity becomes again prevalent, being greater than the
extrapolation of the normal state expression, and giving the evidence for
a reduced carrier scattering in the superconducting state.

The conclusion of our analysis is that the Hall anomaly in YBCO thin films
is the result of a delicate interplay of three contributions to the Hall
conductivity: \textit{(i}) positive quasiparticle vortex-core contribution,
associated with normal-state excitations, which dominates at high fields
($B>3$ T) and increases above the extrapolation from the normal state
below $T_{c}$, indicating reduced quasiparticle scattering in superconductor
state; \textit{(ii}) superconducting contribution (excess Hall effect),
resulting from the vortex flux-flow and superconducting fluctuations, which,
by its negative sign, is connected to the details of the Fermi surface,
and is essential to the sign change occurrence in fields below 3 T; and
\textit{(iii}) pinning contribution, which does not contribute significantly
to the Hall conductivity in the investigated range of magnetic fields and
temperatures, but results in an apparent decrease of the hole-particle
asymmetry at low fields. The pinning contribution eventually leads to the
second sign reversal of the Hall effect in YBCO in very low fields ($B<0.5$
T). Finally, we have found that using the experimental values of $\sigma _{xx}$
in the calculation of the Hall resistivity removes the apparent quantitative
discrepancy between the NE model and the experimental data.

\begin{acknowledgments}
This work was supported by the Austrian Fonds zur F\"{o}rderung der wissenschaftlichen
Forschung (Vienna) and by the NSF Grant No. DMR-0073366 (Rochester). Stimulating
correspondence and discussions with R. Ikeda, A.A. Varlamov, Y. Matsuda
and J. Kolacek, are gratefully acknowledged. We would also like to thank
H. Ebisawa for sending their manuscript prior to publication.
\end{acknowledgments}
\bibliographystyle{apsrev}
\bibliography{Hall03long}

\end{document}